\documentclass[12pt]{article}
\usepackage{amssymb,graphicx}
\usepackage{epstopdf}
\usepackage{amsmath,amsfonts}
\usepackage{epsfig}

\def\d{\partial}
\def\l{\left(}
\def\r{\right)}

\newcommand{\be}{\begin{equation}}
\newcommand{\ee}{\end{equation}}
\newcommand{\bea}{\begin{eqnarray}}
\newcommand{\eea}{\end{eqnarray}}
\newcommand{\bg}{\begin{gather}}
\newcommand{\eg}{\end{gather}}
\newcommand{\bseq}{\begin{subequations}}
\newcommand{\eseq}{\end{subequations}}

\begin{document}
\begin{flushright}
\end{flushright}
\vspace{10pt}
\begin{center}
  {\LARGE \bf On chiral magnetic effect and holography} \\
\vspace{20pt}
V.A. Rubakov\\
\vspace{15pt}
\textit{
Institute for Nuclear Research of
         the Russian Academy of Sciences,\\  60th October Anniversary
  Prospect, 7a, 117312 Moscow, Russia}\\
    \end{center}
    \vspace{5pt}

\begin{abstract}

We point out that there is 
a difference between the behavior of fermionic systems
(and their holographic analogs) in a background axial vector
field, on the one hand, and at finite chiral chemical potential,
on the other. In the former case, the electric current induced
by constant background axial field $A_0$ and magnetic field
${\bf B}$ vanishes, while in the latter it is given by the
anomaly-prescribed formula   ${\bf j} = \frac{\mu_A}{2\pi^2}e^2 N_c {\bf B}$.

\end{abstract}

\section{Introduction and summary}

The chiral magnetic effect 
(CME)~\cite{Kharzeev:2004ey,Fukushima:2008xe},
in its simplest version,
is that
chirally asymmetric quark matter in background magnetic field
${\bf B}$ develops electric current directed along ${\bf B}$.
The ``canonical'' expression for the current in a theory with one
quark flavor of unit electric charge $e=1$ and $N_c$ colors 
is~\cite{Fukushima:2008xe} (see also 
Refs.~\cite{Redlich:1984md,Rubakov:1986am,Diakonov:1991mp,Alekseev:1998ds,Son:2004tq,Newman:2005as} 
for related discussion)
\be
{\bf j} = \frac{\mu_A}{2\pi^2}N_c {\bf B} \; ,
\label{1}
\ee
where $\mu_A$ is the chemical potential to the chiral charge.
There is some debate on whether this result is subject to
strong interaction 
corrections~\cite{Bergman:2008qv,Buividovich:2009wi,Yee:2009vw,Rebhan:2009vc,Gorsky:2010xu,Fukushima:2010zz}, and, in particular,
whether it holds in holographic models of 
QCD~\cite{Yee:2009vw,Rebhan:2009vc,Gorsky:2010xu}.
The purpose of this note is to point out that 
Refs.~\cite{Yee:2009vw,Rebhan:2009vc,Gorsky:2010xu}
study, in fact, quite a different situation than that relevant for CME.
In effect, they consider the electric current induced by the
joint action of the background magnetic field ${\bf B}$
and {\it background temporal
component of an axial vector field} $A_\mu^A$ (the latter field 
couples to
the chiral current). Furthermore, they require that when both
electromagnetic (vector) field $A^V_\mu (x)$ and axial vector
field $A^A_\mu (x)$
are present, 
the theory remains
invariant under electromagnetic gauge transformations.
The latter requirement yields the Bardeen counterterm~\cite{Bardeen}
that contributes to the expression for the electric
current\footnote{Ref.~\cite{Gorsky:2010xu} proceeds by adding a
contribution to the electric current coming from 
the scalar sector of the holographic model,
and arrives at several expressions, one of which coincides with
\eqref{1}. Ref.~\cite{Yee:2010} claims  that the result
\eqref{1} is obtained for canonical
ensemble, while the electric current 
calculated for grand canonical ensemble vanishes.}.

Our observation is twofold. First the chemical potential is {\it not}
the same thing as the static and homogeneous limit of
the background axial vector field. Hence, the requirement of
electromagnetic gauge invariance in background fields
$A^V_\mu(x)$ and $A^A_\mu(x)$ 
is irrelevant; in particular, the Bardeen counterterm
is of no direct significance. Second, the chemical potential
can be introduced to a {\it conserved} quantum number only. In the
CME case, this is a suitably defined conserved axial charge.
Unlike the conserved axial current, which is not invariant under
electromagnetic gauge transformations, the conserved axial charge
is well defined, as it is invariant under spatially localized
electromagnetic gauge transformations~\footnote{We leave aside
genuine non-conservation of chiral charge due to triangle anomaly
involving color gauge fields. The latter is treated separately
in the analysis of CME.}. 

We will see that once our observation is accounted for, the result 
\eqref{1} is back. This must be the case, as the formula \eqref{1}
is a direct consequence of the triangle anomaly, as we argue in the end
of this note (cf. 
Ref.~\cite{Newman:2005as}).

This note is organized as follows. In section \ref{sec:2} we
reproduce the result of Ref.~\cite{Rebhan:2009vc} 
(see also Refs.~\cite{Yee:2009vw,Gorsky:2010xu}) 
in a simple ads/QCD-like 
setup. The reader may safely skip this section over: the result is of 
general nature and is easily understood, as we discuss in the beginning
of section~\ref{sec:3}. We then introduce the chemical potential to the
conserved axial charge and reproduce the result~\eqref{1}.

\section{Background vector and axial vector fields}
\label{sec:2}

To illustrate our points, let us consider the simplest
adS/QCD setup, namely, $U(1)$ gauge theory in 5 dimensions on an
interval $x^5\equiv z \in (0,L)$~\cite{Son:2003et}. 
We will argue in section~\ref{sec:3} that the final result
is, in fact, model-independent.
We begin with the action 
\[
S=\int~d^4x dz \l-\frac{1}{4g^2} F_{MN}F_{MN}
- \kappa \epsilon^{MNPQR} A_M F_{NP}F_{QR} \r
\]
where $M,N=0,1,2,3,5$, $g$ is 5-dimensional coupling and
\[
\kappa = \frac{N_c}{24\pi^2} \; .
\]
The bulk field equation reads
\be
\frac{1}{g^2}\d_N F_{NM} - 3\kappa \epsilon^{MNPQR} F_{NP}F_{QR} = 0\; .
\label{2}
\ee
In particular, for $M=5$ one has
\be
\frac{1}{g^2}\d_\mu F_{\mu 5} - 3\kappa \epsilon^{\mu \nu \lambda \rho} 
F_{\mu \nu}F_{\lambda \rho} = 0\; .
\label{3}
\ee
Even though the behavior of this system in the background
vector and axial vector fields is not directly relevant for CME,
let us discuss this behavior 
to make contact with Refs.~\cite{Yee:2009vw,Rebhan:2009vc,Gorsky:2010xu}. 
To this end, we introduce
the background right and left vector fields
$A_\mu^{R,L} (x^\nu)$,
or, equivalently, background vector and axial fields $A^V_\mu= A_\mu^R
+ A_\mu^L$
and $A^A_\mu=A_\mu^R - A_\mu^L$.
The background fields
are identified with the boundary values of the 5-dimensional field,
\[
A_\mu (x^\nu,0) = A_\mu^L (x^\nu) \;\, \;\;\;\;\;\;\;
A_\mu (x^\nu,L) = A_\mu^R (x^\nu) \; .
\]
The currents are obtained, as usual, by varying the action
with respect to these fields. Provided  the bulk equations
\eqref{2} are satisfied, the expressions for 
 the vector and axial
currents 
are
\begin{align*}
j_\mu &= j_\mu^R + j_\mu^L =
\frac{1}{g^2}\left[ F_{\mu 5} (z=L) - F_{\mu 5}(z=0) \right]
-2\kappa \epsilon^{\mu \nu \lambda \rho} (A_\nu^V F_{\lambda \rho}^A
+  A_\nu^A F_{\lambda \rho}^V )
\\
j_\mu^A &=  j_\mu^R - j_\mu^L = 
\frac{1}{g^2}\left[ F_{\mu 5} (z=L) + F_{\mu 5}(z=0) \right]
-2\kappa \epsilon^{\mu \nu \lambda \rho} (A_\nu^V F_{\lambda \rho}^V
+  A_\nu^A F_{\lambda \rho}^A )
\end{align*}
Note that at this stage,
neither vector nor axial current is invariant under the
electromagnetic gauge transformations acting on $A^V_\mu$.

Making use of the field equation \eqref{3}, one finds
for the divergencies
\[
\d_\mu j_\mu = 2\kappa  
F_{\mu\nu}^A \tilde{F}_{\mu \nu}^V
\; , \;\;\;\;\;\;\;
\d_\mu j_\mu^5 = \kappa  (F_{\mu \nu}^V \tilde{F}_{\mu \nu}^V
+  F_{\mu \nu}^A \tilde{F}_{\mu \nu}^A) \; ,
\]
where $\tilde{F}_{\mu \nu} = \frac{1}{2} 
\epsilon^{\mu \nu \lambda \rho}  F_{\lambda \rho}$.
This is the same result as in 
Ref.~\cite{Rebhan:2009vc}.
Aiming at restoring the conservation of vector current,
one adds the Bardeen counterterm into the action,
\be
S_B = 2\kappa \int~d^4x~ \epsilon^{\mu \nu \lambda \rho}  
A_\mu^A A_\nu^V F^V_{\lambda \rho}
\label{may6-1}
\ee
The corresponding Bardeen currents are
\[
j_{\mu \, B}  = -4 \kappa  \epsilon^{\mu \nu \lambda \rho} A_\nu^A 
F_{\lambda \rho}^V + 2\kappa   \epsilon^{\mu \nu \lambda \rho} 
A_\nu^V F_{\lambda \rho}^A
\;, \;\;\;\;\;\;\;
j_{\mu, B}^5  = 2\kappa 
\epsilon^{\mu \nu \lambda \rho} A_\nu^V F_{\lambda \rho}^V \; .
\]
With these terms included, the currents become
\begin{align}
J_\mu & = j_\mu + j_{\mu \, B} =
\frac{1}{g^2} \left[F_{\mu 5} (z=L) - F_{\mu 5}(z=0) \right]
-  6 \kappa
\epsilon^{\mu \nu \lambda \rho} A_\nu^A F_{\lambda \rho}^V
\label{6}
\\
J_\mu^5 & = j_\mu^5 + j_{\mu \, B}^5 =
\frac{1}{g^2} \left[F_{\mu 5} (z=L) + F_{\mu 5}(z=0) \right]
-  2 \kappa
\epsilon^{\mu \nu \lambda \rho} A_\nu^A F_{\lambda \rho}^A
\end{align}
The gauge invariance of the currents under the electromagnetic
gauge transformations is now restored, while the divergencies are
\be
\d_\mu J_\mu = 0 \; ,\;\;\;\;\;\;\;\;
\d_\mu J_\mu^5 =  3\kappa  
F_{\mu\nu}^V \tilde{F}_{\mu \nu}^V +  \kappa  
F_{\mu\nu}^A \tilde{F}_{\mu \nu}^A
\label{may6-2}
\ee
This is the standard result.

Finally, let us consider static background with non-vanishing
vector-potentials
$A_i^V = A_i^V ({\bf x})$, $A_0^A = \mbox{const}$, 
and constant magnetic field $B_i = \frac{1}{2} \epsilon^{ijk}F^V_{jk}$.
The field equations for static fields $A_\mu = A_\mu ({\bf x},z)$,
$A_5=0$ read
\begin{align}
\frac{1}{g^2} \d_5^2 A_i - 12 \kappa \epsilon^{ijk} \d_5 A_0 F_{jk} &= 0
\label{5}
\\
\frac{1}{g^2} \d_5^2 A_0 + 12 \kappa \epsilon^{ijk} \d_5 A_i F_{jk} &= 0
\label{4}
\\
\d_5 \d_i A_i & =0
\end{align}
This relevant solution to system has the form
\[
A_i ({\bf x}, z) = \frac{1}{2} A_i^V ({\bf x}) + C_i (z) \;,\;\;\;
\;\;\;\; A_0 = A_0 (z)
\]
with boundary conditions $C_i(L) = C_i(0) =0$,
$A_0 (L) = \frac{1}{2} A_0^A$,
$A_0 (0) = - \frac{1}{2} A_0^A$. The function $A_0 (z)$ is antisymmetric
with respect to the point $L/2$, and Eq.~\eqref{4} shows that $C_i(z)$
is symmetric. We derive from Eq.~\eqref{5} that 
\[
\frac{1}{g^2}F_{i5} = - \frac{1}{g^2} \d_5 C_i = - 12 \kappa A_0 B_i
\]
and hence
\[
\frac{1}{g^2} \left[F_{i5}(L) - F_{i5}(0) \right] = 
-12 \kappa B_i \left[A_0 (L) - A_0 (0) \right] =  
-12 \kappa B_i A_0^A \; .
\]
This term cancels out the second term in the expression
\eqref{6} for the electric current $J_i$, so the current
vanishes in the background field configuration we consider,
\[
J_i =0 \; .
\]
This is the result obtained in 
Ref.~\cite{Rebhan:2009vc}
(see also Refs.~\cite{Yee:2009vw,Gorsky:2010xu}).

\section{Axial chemical potential}
\label{sec:3}

The fact that the electromagnetic current vanishes
in the backround of constant axial vector potential $A_0^A$ 
and magnetic field ${\bf B}$, at least to the
first order in ${\bf B}$, 
is of general nature. To see this,
consider the effective action for the fields $A^V_\mu$ and $A^A_\mu$
obtained by integrating out the dynamical degrees of freedom.
In terms of it, the electric current is
\[
J_i = \frac{\delta S_{eff}[A^V,A^A]}{\delta A^V_i} \; .
\]
If it did not vanish in our background, and 
had the form
${\bf J} \propto A_0^A \cdot {\bf B}$, then the lowest derivative term
in the effective action would  
have precisely the form of the Bardeen action,
\be
S_{eff} [A^V, A^A]= 
\mbox{const} \cdot \int~d^4x~ \epsilon^{\mu \nu \lambda \rho}  
A_\mu^A A_\nu^V F^V_{\lambda \rho}
\label{may6-3}
\ee
However, the theory, and hence the effective action,  is invariant
under electromagnetic gauge transformations, so such a term
cannot be generated.

Let us now switch off the axial vector field $A_\mu^A$ and introduce
instead finite axial chemical potential $\mu_A$.
In the first place, the chemical potential is constant in space and
time, so the constraints coming from the requirement of the
electromagnetic gauge invariance are relaxed. Second, the chemical 
potential can be introduced to a conserved quantity only.
In other words, a quantity well defined for a microcanonical
ensemble is a quantum number that does not change  when
other parameters (like background fields) vary.
Precisely because of the anomaly \eqref{may6-2}, the integral of
$J^5_0$ over space is not conserved. The conserved chiral charge
is
\[
Q^5 = \int~d^3x~J_0^5 - 3 \kappa \int~d^3x~\epsilon^{ijk} A_i^V F_{jk}^V
\]
Since $J_0$ is gauge invariant, this chiral charge is invariant
under the electromagnetic gauge transformations. 
Upon adding the chemical potential, the (Euclidean) action 
of the theory becomes
\be
  S[\mu] = S - \mu_A \int~dx^0~Q^5 =
\left( S-\int~d^4x \mu_A J_0^5\right) + 3 \kappa \mu_A \int~d^4x~
\epsilon^{ijk} A_i^V F_{jk}^V
\label{may6-4}
\ee
where $S$ is the original action of the theory. The dynamical degrees
of freedom enter only the term in parenthesis, which is invariant
under electromagnetic gauge transformations even for $\mu_A$
varying in space and time. In other words, when considering the dynamics
induced {\it by this term}, one {\it can} treat $\mu_A$ as the static
and homogeneous axial vector field (in the model of section~\ref{sec:2}
this is precisely the dynamics studied there). According to the above
argument, this dynamics does not induce the term 
\eqref{may6-3} in the effective action. Hence, the lowest derivative
term in the effective
action is simply given by the last term in
\eqref{may6-4} (cf. Ref.~\cite{Rubakov:1986am}),
\be
  S_{eff} = 3 \kappa \mu_A \int~d^4x~
\epsilon^{ijk} A_i^V F_{jk}^V
\label{may6-5}
\ee
By varying this effective action with respect to $A_i$, one
arrives at the result \eqref{1}.

To conclude, any model with correct anomaly structure yields the effective
action \eqref{may6-5}, and hence the exression \eqref{1} for the electric
current induced in chirally asymmteric matter.

\vspace{0.3cm}

The author is indebted to A.~Gorsky,
D.~Kharzeev, D.T.~Son and M.~Stephanov for
useful comments. This work
has been supported in part by Russian Foundation for Basic
Research grant 08-02-00473.

\end{document}